\documentclass{aastex}

\shorttitle{Optical Monitoring of S5 0716+714}
\shortauthors{Xiong et
al.}

\begin{document}
\title{Optical quasi-periodic oscillation of the BL Lacertae object S5 0716+714 during the faint state}

\author{Shanwei Hong,}
\affil{Yunnan Observatories,
Chinese Academy of Sciences, 396 Yangfangwang, Guandu District, Kunming, 650216, P. R. China}
\affil{University of Chinese Academy of Sciences, Beijing 100049, China}
\affil{School of Applied Technology, Lijiang Teachers College, Lijiang 674199, China}

\author{Dingrong Xiong and Jinming Bai}
\affil{Yunnan Observatories,
Chinese Academy of Sciences, 396 Yangfangwang, Guandu District, Kunming, 650216, P. R. China}
\affil{Center for Astronomical Mega-Science, Chinese Academy of Sciences, 20A Datun Road, Chaoyang District, Beijing, 100012, P. R. China}
\affil{Key Laboratory for the Structure and Evolution of Celestial
Objects, Chinese Academy of Sciences, 396 Yangfangwang, Guandu District, Kunming, 650216, P. R. China}\email{xiongdingrong@ynao.ac.cn}

\begin{abstract}
In order to search for the evidence of quasi-periodic oscillation (QPO) in blazar, multicolor optical observations of the BL Lacertae object S5 0716+714 were performed from 2005 to 2012. For $I$ band on March 19 2010 with low variability amplitude and low flux level, the same quasi-periodic oscillation $\simeq50$ minutes with 99\% significance levels is confirmed by ZDCF method, Lomb-Scargle method, REDFIT and fitting sinusoidal curves. The observed QPO is likely to be explained by accretion disk variability. If the observed QPO indicates an innermost stable orbital period from the accretion disk, the QPO $\simeq50$ min corresponds to a black hole mass of $5.03\times10^6 M_\odot$ for a non-rotating Schwarzschild black hole and $3.2\times10^7 M_\odot$ for a maximally rotating Kerr black hole.
\end{abstract}


\section{INTRODUCTION}

Blazar is an extreme subclass of active galactic nuclei (AGNs), whose jets point in the direction of the observer (Urry \& Padovani 1995; Ghisellini \& Tavecchio 2015). On the basis of the differences of equivalent width of the emission lines, blazar is often divided into two subclasses: BL Lacertae (BL Lac) object and flat spectrum radio quasar (FSRQ). FSRQ has strong emission lines, while BL Lac object has very weak or non-existent emission lines (Urry \& Padovani 1995; Xiong et al. 2017; Qin et al. 2017). Blazar shows variability over the entire electromagnetic spectrum (Falomo et al. 2014). According to different spans of timescales, its variability can be classed into three types. Flux/Brightness changes from a few minutes to less than a day are called as intraday variability (IDV) or microvariability, from days to months as short-term variability and from months to years as long-term variability (Wagner \& Witzel 1995; Dai et al. 2015; Hong et al. 2017). For optical IDV of blazars, detection of periodic or quasi-periodic
oscillations (QPOs) would be a strong evidence for the presence of a single dominant orbiting hot-spot on the accretion disk, or accretion disk pulsation (Gupta et al. 2009). QPQs in less than one day timescale were detected just for a few blazars (Espaillat et al. 2008; Gupta et al. 2009; Rani et al. 2010; Lachowicz et al. 2009). Signals of QPOs still need to further be confirmed for them.

S5 0716+714 is one of the best studied sources across the electromagnetic spectrum due to high variability with very high duty cycle of IDV (Hong et al. 2017 and therein references). It is classified as an intermediate-synchrotron-peaked BL Lac object, which has featureless optical continuum (Abdo et al. 2010; Gupta et al. 2008). Making use of the host galaxy as a standard candle, Nilsson et al. (2008) presented the redshift value of $0.31\pm0.08$ for the source. Danforth et al. (2013) constrained the redshift range from 0.2315 to 0.3407 by using intervening absorption systems. It is not possible to determine the black hole mass by using these methods that require spectroscopy because the optical spectrum of S5 0716+714 is a featureless continuum (Gupta et al. 2009). If the observed periodic variability timescale indicates an
innermost stable orbital period from the accretion disk, then periodic variability timescales can estimate/limit its black hole mass (e.g., Fan et al. 2014; Gupta et al. 2009; Dai et al. 2015). However, such periodic variability timescales are not easily confirmed because periodic variability is rarely found and many possible variability timescales are not related with innermost stable orbital period from the accretion disk. From the beginning
of 1995 to the beginning of 2007, five outbursts from the source suggest a possible period of long-term variability of $\sim3.0\pm0.3$ years (Raiteri et al. 2003; Foschini et al. 2006; Gupta et al. 2008). A possible 10 day period was observed simultaneously in the $BVRI$ bands (Dai et al. 2015). Quirrenbach et al. (1991) found that the blazar has a possible QPO on the timescale of $\sim1$ day. Stalin et al. (2009) presented the results of intra-night optical monitoring of S5 0716+714 who gave evidences for possible quasi-periods ranging
from 0.9 to 4.3 h. Gupta et al. (2009) found high probabilities (from 95\% to $>$99\%) quasi-periodic IDV timescales between $\sim$25 and $\sim73$ minutes (min), which is the first good evidence for quasi-periodic components in the optical intra-day variable light curves of any blazars. Using several different techniques, Rani et al. (2010) discovered an approximately 15 min quasi-period at a $>3\sigma$ confidence level for S5 0716+714.

In order to search for the evidence of QPOs, we observed the target from 2005 to 2012. Section 2 describes the observations and data reduction. The results are reported in Section 3. In Section 4, we present discussion and conclusion.

\section{OBSERVATIONS AND DATA REDUCTION}

In order to explore mechanism of variability, we monitored many radio loud AGNs using the 2.4 m, 1.02 m and 60 cm optical telescopes located at the Yunnan Astronomical Observatories (YAO) of Chinese Academy of Sciences of China from 2005 to 2012 (e.g., Liao et al. 2014; Xiong et al. 2016, 2017; Wang et al. 2017; Guo et al. 2017; Hong et al. 2017). The S5 0716+714 is one of them. Currently, some radio loud AGNs are still observed. The standard Johnson UBV and Cousins RI filters were equipped for the three telescopes. After 2012, the filters of 60 cm optical telescope were changed into ugri of SDSS. Further details about the three telescopes are seen in Liao et al. (2014), Xiong et al. (2017) and Hong et al. (2017). In this work, S5 0716+714 is observed with 2.4 m and 1.02 m telescopes. The observation log is given in Table 1. Our photometry observations were performed by two different exposure modes: the first was that all of the observations were completed for the same optical band and
then moved to the next band (AAA...BBB...); the second was a cyclic mode among different bands (ABCDABCD...). Table 1 shows that most of nights have time resolution less than 5 minutes and time span less than 1 hour. The exposure times from 0.09 to 6 minutes were chosen according to different seeing and brightness of the source. During the observations, about 5 seconds of exposure is observed with 2.4 m telescope. The photometric errors from IRAF are from 0.002 to 0.003 in about 5 seconds of exposure. According to definition of $S/N={\rm Flux/Error}$, the $S/N$ is approximately equal to the reciprocal of errors ($1.0857*{\rm Error/Flux}$) from IRAF. Therefore, the values of $S/N$ in about 5 seconds of exposure correspond to $333-500$. Goyal et al. (2013) found that IRAF produces photometric errors that are too small by factors of $\sim1.5$. If so, the actual values of $S/N$ should be $222-333$. For our original data, Liao et al. (2014) used partial data ($\sim45\%$) to study multi-wavelength variability properties of the source. However, they only made use of mean magnitude on each day and their scientific goals were different from ours. So we rehandled the data and kept them in our present work (see Table 1). The aperture photometry was performed using the DAOPHOT task of IRAF software package after flat-field and bias were corrected. We used different aperture radii (1.5$\times$FWHM, 1.7$\times$FWHM, 2$\times$FWHM, 2.5$\times$FWHM, 3$\times$FWHM) to carry out aperture photometry. Considering the best signal-to-noise ratio, we chose aperture radius of 2$\times$FWHM. Two comparison stars in the same frame, stars 5 and
6 in the finding chart of Villata et al. (1998), were adopted because they are close to S5 0716+714 in the frame and have similar apparent magnitudes with the source (Dai et al. 2015). The magnitude of S5 0716+714 was calculated as the average of that derived with respect to comparison stars 5 and 6 (Zhang et al. 2004, 2008; Fan et al. 2014; Bai et al. 1998). The rms errors from comparison stars 5 and 6 were used to estimate the observational uncertainty (Fan et al. 2014; Xiong et al. 2016). The photometric results of the observations are given in Table 2--5, including 91 observational nights.

\section{RESULTS} \label{bozomath}

We have employed two statistical analysis techniques ($F$ test and the one-way analysis of variance: ANOVA) to quantify the IDV/microvariation of the BL Lac object (e.g., de Diego 2010; Goyal et al. 2012; Hu et al. 2014; Agarwal \& Gupta 2015; Dai et al. 2015; Xiong et al. 2016, 2017; Hong et al. 2017). The specific formulas and variability criterion about the two statistical techniques are seen in these references (e.g., de Diego 2010; Goyal et al. 2012; Hu et al. 2014; Agarwal \& Gupta 2015; Dai et al. 2015; Xiong et al. 2016, 2017; Hong et al. 2017). The BL Lac object is considered as variability (V) if the light curve satisfies the two criteria of $F$-test and ANOVA, probably variable (PV) if only one of the above two criteria is satisfied, and non-variable (N) if none of the criteria are met. We only analyze the nights with $N\geq15$. Nine nights show IDV. The duty cycle of variability (Romero et al. 1999; Stalin et al. 2009) is $19.57\%$ and $31.34\%$ for V and V+PV cases respectively. The analysis results of IDV are given in Table 6. The Fig. 1 shows the light curves detected as IDV. After that, we use the z-transformed discrete correlation function (ZDCF; Alexander 1997) to analyze possible QPOs. The descriptions of ZDCF method can be found in Xiong et al. (2017). For the autocorrelation function (ACF) calculated by ZDCF method, the presence of strong peaks in autocorrelation apart from the one near zero indicates a possible QPO (e.g., Stalin et al. 2009). Among these nights detected as IDV, $I$-band light curve on March 19 2010 shows a possible QPO (upper panel in Fig. 2). The timescale of QPO in highest point of the peak is 0.031 day (44.64 min). Making use of a (error-weighted) least-squares procedure, we use a fifth-order polynomial to fit the ACF and consider ACF($\tau=0$)=1 (Giveon et al. 1999; Xiong et al. 2017). The fitting result shows a possible QPO$=$0.035 day (50.4 min). The Lomb-Scargle method (Lomb 1976 and Scargle 1982) is commonly used to detect periodicity in the spectrum of time series including unevenly sampled time series (Li et al. 2016; Xiong et al. 2017). We also employ the Lomb-Scargle method to analyze the possible QPO on March 19 2010 (middle panel in Fig. 2). For a power level $z$ within Lomb-Scargle method, the False Alarm Probability (FAP) is calculated by $p(>z)\approx N\cdot {\rm exp}(-z)$, where $N$ is the number of data point (Scargle 1982; Press et al. 1994; Li et al. 2016). The results of Lomb-Scargle method show a possible QPO$=$0.036 day (51.84 min) at 0.01 FAP levels. For the periodicity analysis of AGNs, frequency-dependent red noise needs to be considered (Xiong et al. 2017; Fan et al. 2014; Sandrinelli et al. 2016). Schulz \& Mudelsee (2002) presented a computer program (REDFIT) estimating red-noise spectra directly from unevenly spaced time series by fitting a first order autoregressive (AR1) process. The results indicate that the peak ($\frac{1}{1.05}$ hour $\simeq$ 57 min) in the spectrum of a time series is significant against the red-noise background from an AR1 process, with 99\% $\chi^2$ significance levels (bottom panel in Fig. 2). Moreover, sinusoidal curves are applied to fit the data on this night. However, the light curve on this night can not be fitted by sinusoidal curves because there is a continuous dimming trend in the light curve of the night. We use error-weighted linear regression analysis to fit the continuous dimming trend. The continuous dimming trend is subtracted from the original light curve (Fig. 3). For $I$-band data on this night, the light curve is not continuous because of the deficiency of a few data points. During the fitting processes, we assume that the data points have average high or low magnitudes. From Fig. 3, the periods of fitting sinusoidal curves are 0.03 day (43 min) for average low magnitudes and 0.034 day (49 min) for average high magnitudes respectively. Therefore, above results indicate that a possible QPO with timescales from 43 to 57 min (mean value $\simeq50$ min) is detected, i.e., the same QPO $\simeq50$ min with 99\% significance levels is confirmed by ZDCF method, Lomb-Scargle method, REDFIT and fitting sinusoidal curves.

\section{DISCUSSION AND CONCLUSION} \label{bozomath}

Multicolor optical observations of the BL Lac object S5 0716+714 were performed from 2005 to 2012. Making use of two statistical methods, we find that nine nights show IDV. Previous results found that the BL Lac object has a high duty cycle (e.g., Agarwal et al. 2016; Hu et al. 2014; Chandra et al. 2011; Stalin et al. 2009; Hong et al. 2017). The duty cycle of variability from our results is 19.57\% and 31.34\% for V and V+PV cases respectively. Our observational results are inconsistent with previous results. However, it should be noticed that most of time spans on per night are less than one hour for our observations. Extensive IDV studies revealed that the occurrences of IDV in a blazar is related with observational time spans (Gupta \& Joshi 2005; Rani et al. 2011). So, the conclusion on the low duty cycle of this source is solely due to very poor time coverage on most of the nights.

For $I$ band on March 19 2010, the same QPO $\simeq50$ min with 99\% significance levels is confirmed by ZDCF method, Lomb-Scargle method, REDFIT and fitting sinusoidal curves. This is a new result which further confirms the result of Gupta et al. (2009). The Amp is $4.29\%$ and the magnitude is $13^{\rm m}.43$ on this night that has low variability amplitude and flux level (Dai et al. 2015; Hong et al. 2017). For IDV of blazar, there are two main mechanisms (intrinsic and extrinsic origins) to explain it (e.g., Xiong et al. 2017). Intrinsic mechanisms are related with relativistic jet and accretion disk. Extrinsic origins include interstellar scintillation and gravitational microlensing. But extrinsic mechanisms can not explain the origin of the IDV in the optical band (Heeschen et al. 1987; Agarwal \& Gupta 2015). The IDV/short variability is often explained by the shock-in-jet model (Marscher \& Gear 1985; Hu et al. 2014; Dai et al. 2015; Xiong et al. 2016, 2017; Hong et al. 2017). The changes of relativistic jet always dominate the IDV for blazars in a high state while the origin of IDV can be explained by accretion disk variability in a low state (Xiong et al. 2016). Therefore, the origin of IDV on March 19 2010 is likely to be explained by accretion disk variability. The short optical QPO can be explained by  hot spots or some other non-axisymmetric phenomenon related to the orbital motions (Gupta et al. 2009; Rani et al. 2010). If the observed QPO indicates an innermost stable orbital period from the accretion disk, the black hole masses can be estimated as
\begin{equation}
\frac{M}{M_\odot}=\frac{3.23\times10^4P}{(r^{3/2}+a)(1+z)},
\end{equation}
where $r$ is innermost stable orbital radius, $a$ is angular momentum parameter, $P$ is innermost stable orbital period in units of seconds and $z$ is the redshift (Gupta et al. 2009). The QPO $\simeq50$ min corresponds to a black hole mass of $5.03\times10^6 M_\odot$ for a non-rotating Schwarzschild black hole and $3.2\times10^7 M_\odot$ for a maximally rotating Kerr black hole. If the period of the QPO is more than innermost stable orbital period, then the actual black hole mass is less than above value. However, this result cannot totally exclude that the QPO originates from a relativistic shock propagating down a helical jet or turbulence behind a shock propagating down a jet (see Rani et al. 2010 for detail). In addition, during fitting sinusoidal curves, we find that there is a continuous dimming trend in the light curve of the night besides QPO component, which suggests that the optical emission includes different components.

\begin{acknowledgements}
We sincerely thank the referee for valuable comments and
suggestions. We acknowledge the support of the staff of the
Lijiang 2.4m and Kunming 1m telescopes. Funding for the two telescopes has been provided by the Chinese Academy of Sciences and the Peoples Government of Yunnan Province. This work is financially supported by the National Nature Science Foundation of China (11433004, 11133006, 11673057, 11361140347, 11703078), the Key Research Program of the Chinese Academy of Sciences (grant No. KJZD-EW-M06), the Strategic Priority Research Program ``The emergence of Cosmological Structures'' of the Chinese Academy of Sciences (grant No. XDB09000000), the Chinese Western Young Scholars Program
and the ¡°Light of West China¡± Program provided by CAS (Y7XB018001, Y5XB091001), the science and technology project for youth of Yunnan of China (2014FD059).
\end{acknowledgements}

\begin{figure}
\begin{center}
\includegraphics[width=17cm,height=20cm]{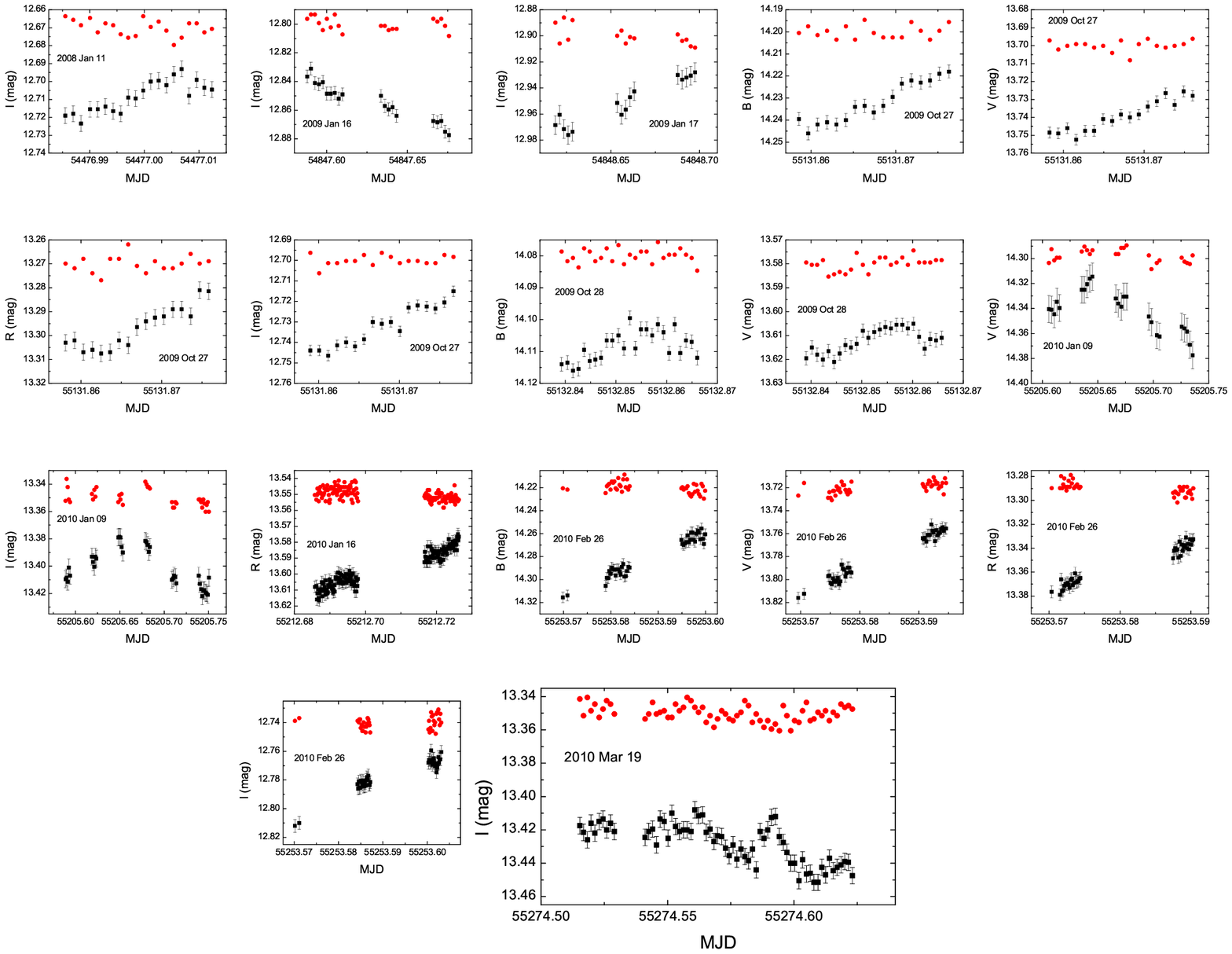}
\caption{Optical IDV light curves. The black
squares stand for the light curves of the source, and the red
circles for the values of $S_x$ (Hu et al. 2014) which can quantify the accuracy of variability.}
\end{center}
\end{figure}

\begin{figure}
\begin{center}
\includegraphics[angle=0,width=0.6\textwidth]{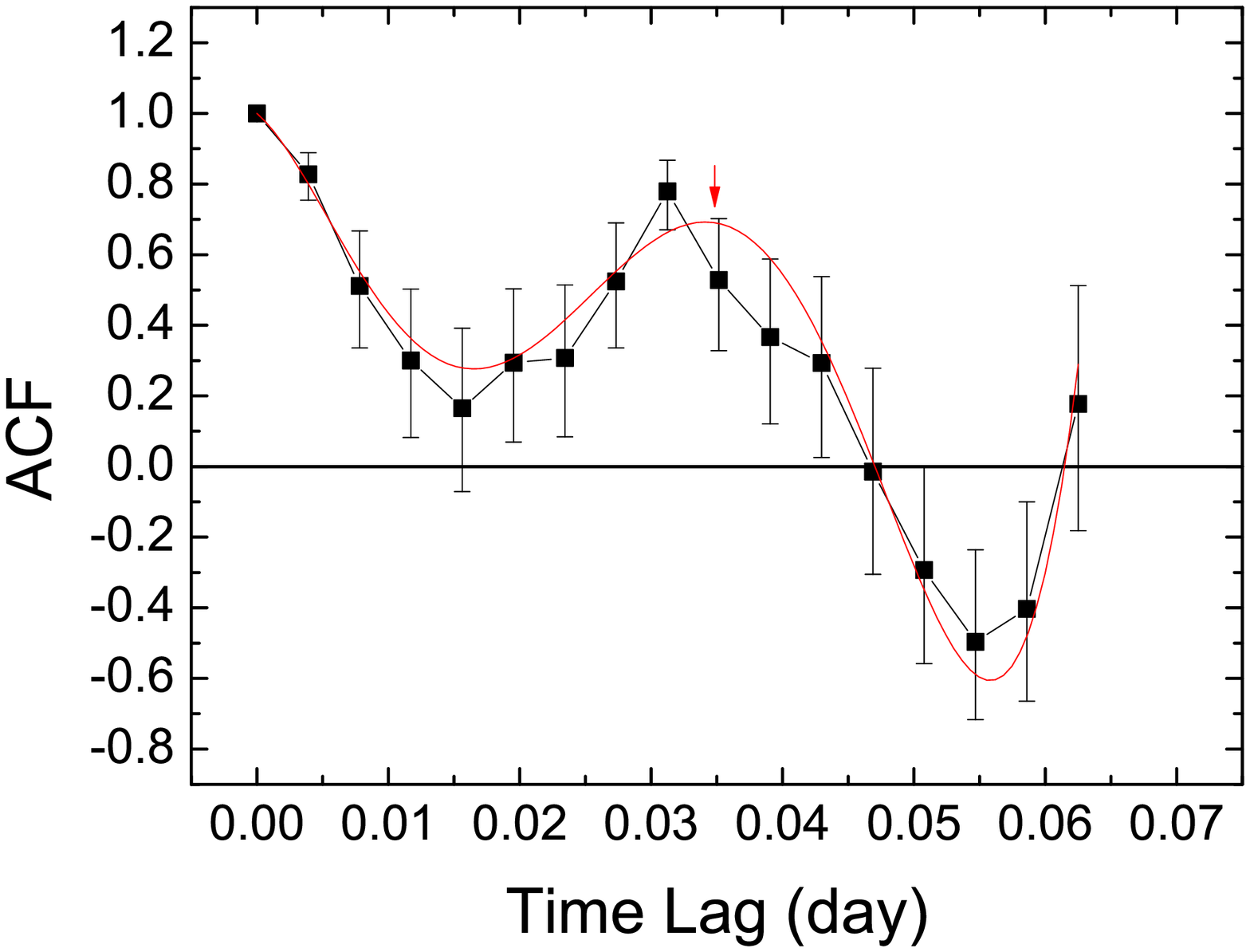}
\includegraphics[angle=0,width=0.6\textwidth]{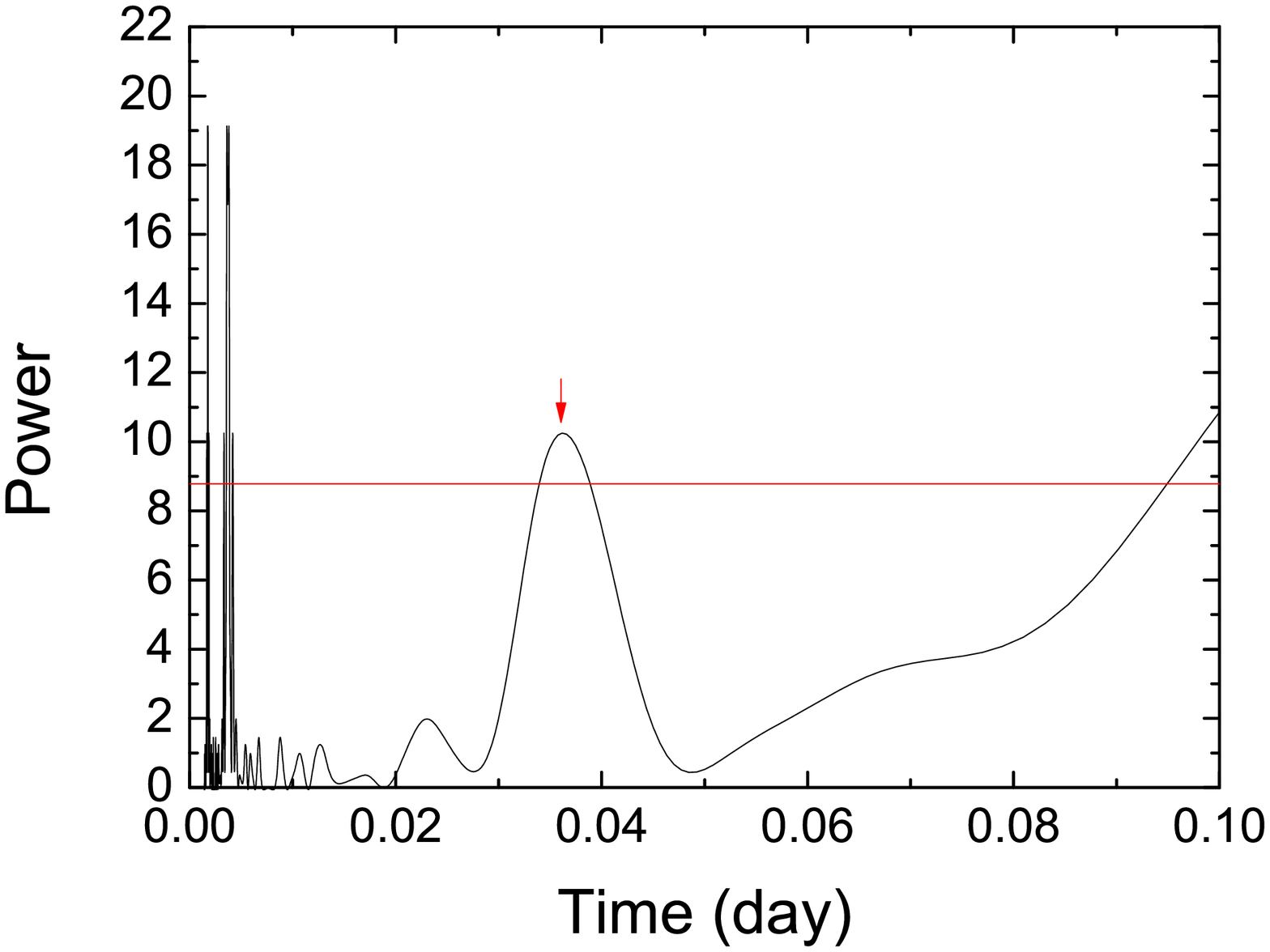}
\includegraphics[angle=0,width=0.6\textwidth]{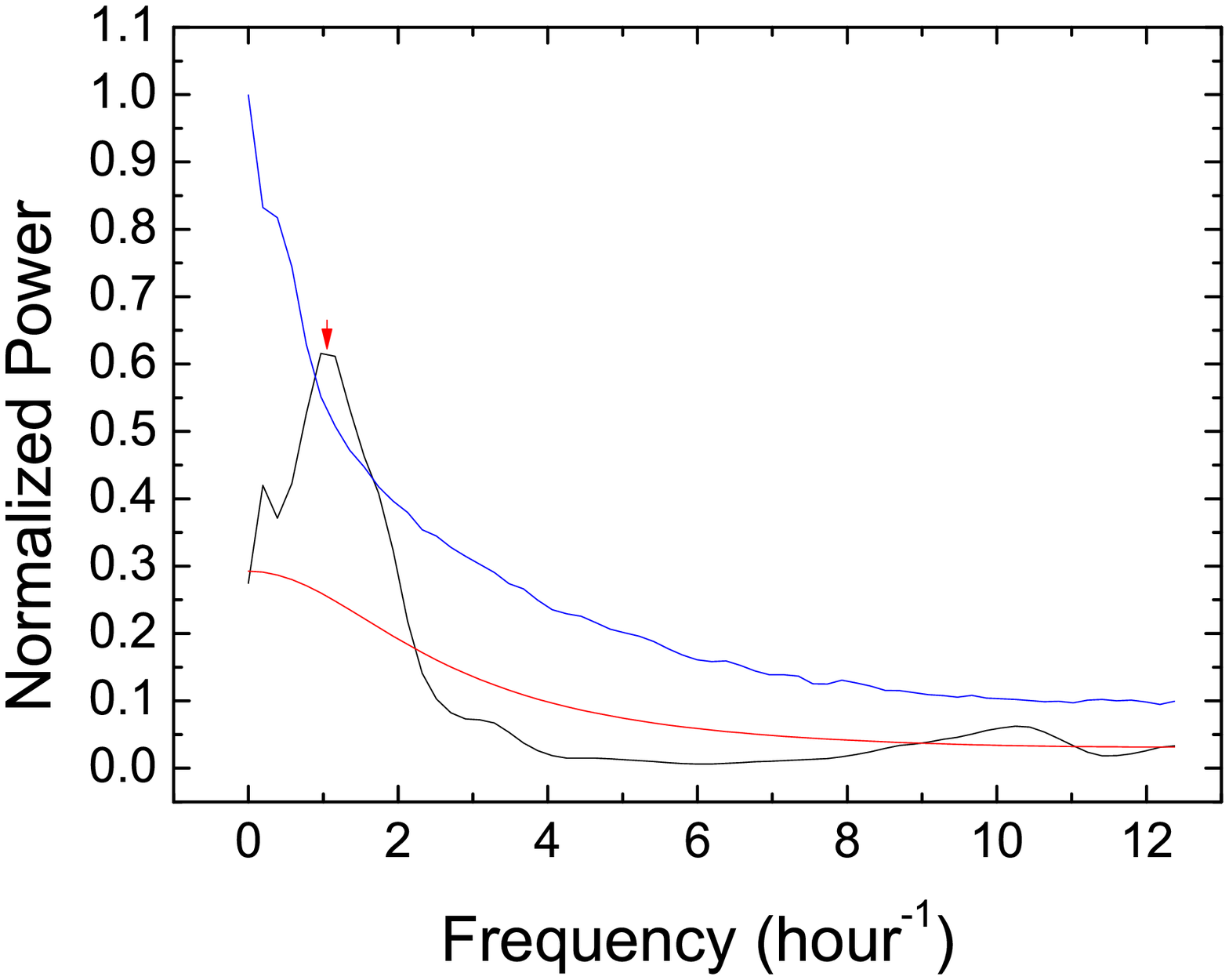}
\caption{{\scriptsize The results of periodicity analysis for $I$-band light curve on March 19 2010. The upper, middle and bottom panels are results of ZDCF method, Lomb-Scargle method and REDFIT respectively. For the upper panel, the red line stands for a fifth-order polynomial least-squares fit. For the middle panel, the red line indicates 0.01 FAP levels. For the bottom panel, the black line is bias-corrected power spectra, the red line is theoretical red-noise spectrum and the blue line is 99\% $\chi^2$ significance levels. The red arrow in each panel refers to a possible signal of QPO.} \label{fig3}}
\end{center}
\end{figure}

\begin{figure}
\begin{center}
\includegraphics[angle=0,width=0.7\textwidth]{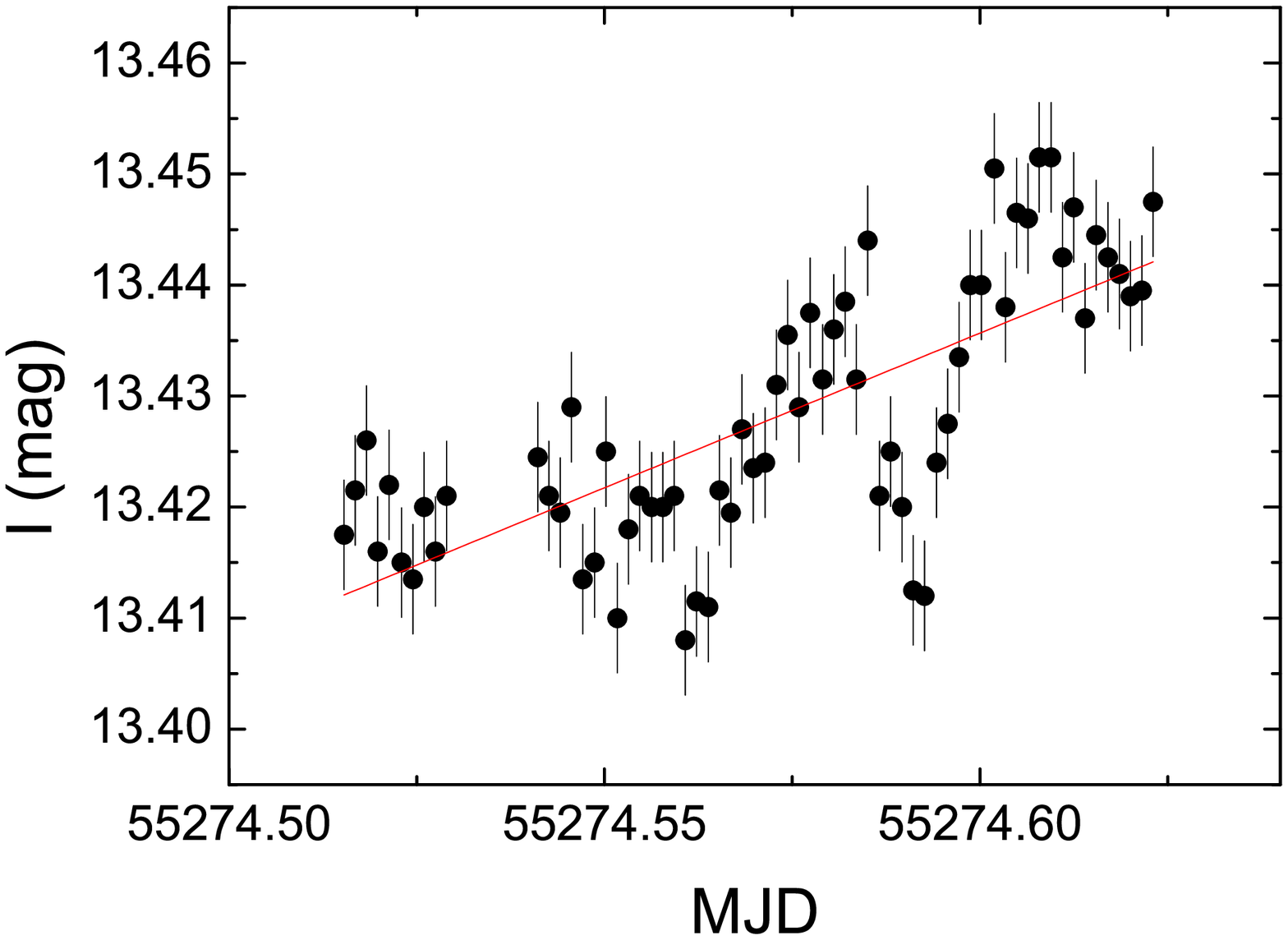}
\includegraphics[angle=0,width=0.7\textwidth]{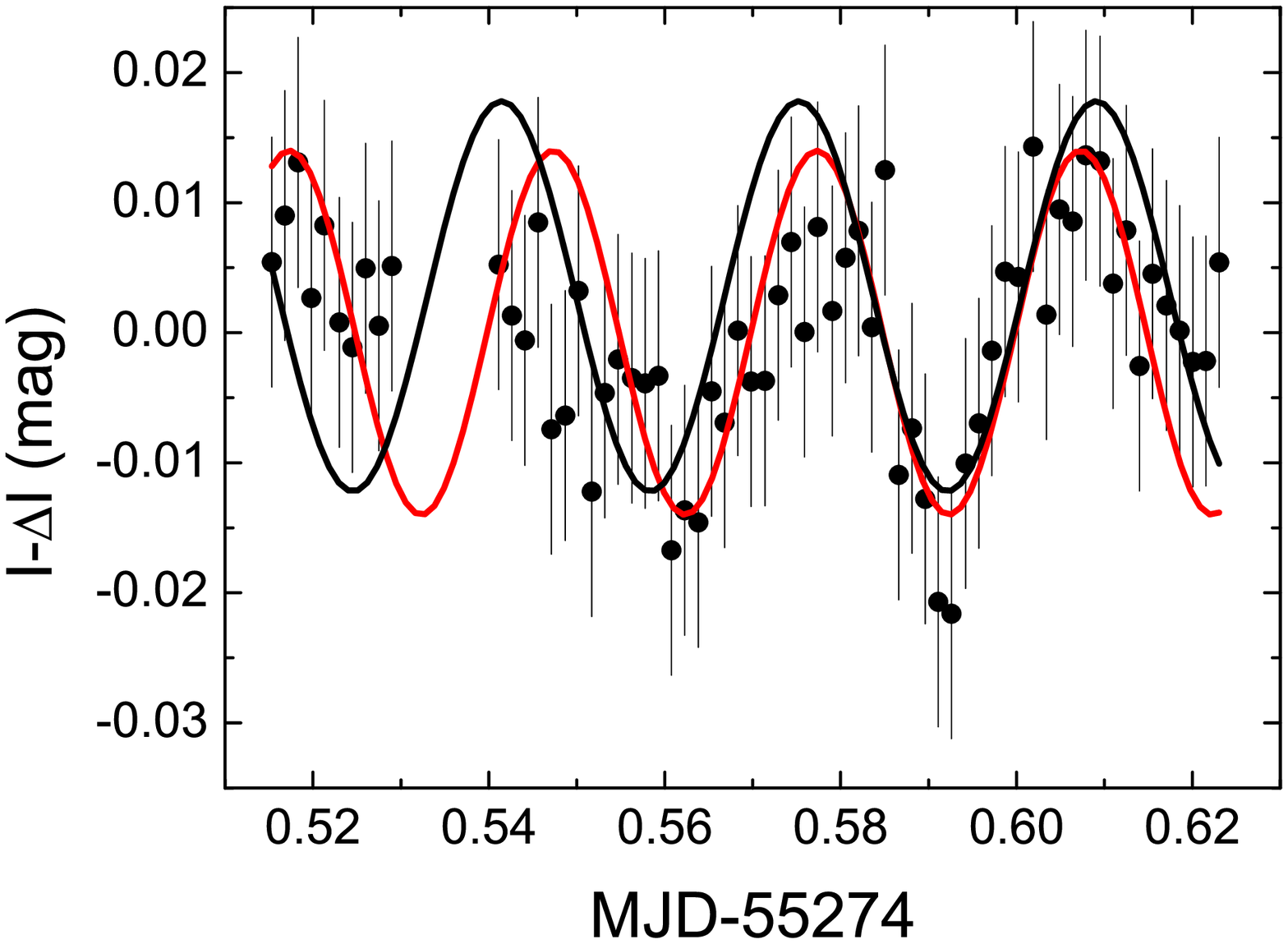}
\caption{Light curves on March 19 2010. For first panel, the red line is the result of error-weighted linear regression analysis. The continuous dimming trend is subtracted from the original light curve for the bottom panel in which the two solid lines stand for sinusoids with different fitting parameters.  \label{fig3}}
\end{center}
\end{figure}

\clearpage

\begin{deluxetable}{ccccr}
\tablecaption{Observation Log of S5 0716+714\label{tbl-3}} \tablewidth{0pt}
\tablehead{\colhead{Date} &\colhead{Bands} & \colhead{Number of data points} &
\colhead{Time spans(h)} & \colhead{Time resolutions(min)}}
\startdata
20050405	&	B,V,R,I	&	2,2,2,2	&	0.25	&	13	\\
20050406$^\dag$	&	B,V,R,I	&	9,10,10,10	&	1.72,1.93,1.94,1.96	&	12	\\
20060105	&	B,V,R,I	&	6,6,6,7	&	0.32,0.3,0.3,0.25	&	4,3.5,3,2.4	\\
20060109	&	B,V,R,I	&	4,4,4,4	&	0.22	&	4,3,2.5,2.5	\\
20060110	&	B,V,R,I	&	7,8,8,8	&	0.83	&	3.5,2.5,2,2	\\
20060111	&	B,V,R,I	&	8,8,8,8	&	0.8	&	3,2,2,3	\\
20061028	&	V,I	&	13,21	&	1.18,1.97	&	3.6,1.5	\\
20061128	&	V,I	&	32,31	&	1.93,1.82	&	2.1,1.5 	\\
20061129	&	B,V,I	&	12,12,12	&	1.12,1.04,0.95	&	3.5,2.5,1.5	\\
20070124	&	V,I	&	4,6	&	0.11,0.17	&	2.1,1.5	\\
20070128	&	I	&	16	&	1.54	&	2	\\
20070327	&	V,I	&	20,20	&	0.82,0.75	&	1.7,1.2	\\
20070328	&	B,V,I	&	15,15,15	&	2.6,2.5,2.3	&	5.5,4.6,2.3	\\
20070329	&	B,V,I	&	10,10,10	&	1.38,1.25,1.13	&	5.5,3.8,2.2	\\
20070330	&	B,V,I	&	10,10,10	&	1.44,1.34,1.18	&	5.5,3.8,2.2	\\
20070415	&	V,I	&	4,5	&	0.25,0.31	&	4.8,2.3	\\
20070416	&	I	&	15	&	0.52	&	2.2	\\
20070422	&	I	&	6	&	0.41	&	1.5	\\
20070423	&	B,I	&	4,5	&	0.43,0.17	&	5.5,2.2	\\
20070424	&	B,V,I	&	5,5,5	&	0.27,0.21,0.21	&	4,3,3	\\
20071027	&	I	&	4	&	0.23	&	4.5	\\
20071028	&	I	&	4	&	0.13	&	2.5	\\
20071120	&	I	&	11	&	0.63	&	2	\\
20071121	&	I	&	12	&	0.46	&	2.5	\\
20080110	&	I	&	20	&	0.65	&	2	\\
20080111	&	R,I	&	10,20	&	0.46,0.64	&	3,2	\\
20080112	&	R,I	&	10,10	&	0.46,0.34	&	3,2.2	\\
20080922$^*$	&	R,I	&	10,10	&	0.49,0.46	&	2.2,1.7	\\
20080924$^*$	&	I	&	10	&	0.74	&	3.9	\\
20081002$^*$	&	I	&	23	&	1.07	&	2.8	\\
20090116$^*$	&	R,I	&	14,20	&	1.93,2.11	&	5.2,3.5	\\
20090117$^*$	&	R,I	&	10,15	&	1,18,1.89	&	5.2,3.5	\\
20090118$^*$	&	R,I	&	10,9	&	1.11,1	&	5.2,3.5	\\
20090119$^*$	&	B,V,R,I	&	8,8,8,7	&	0.13,0.14,0.12,0.11	&	1.1,1.1,1,1	\\
20090120$^*$	&	R,I	&	15,15	&	1.99,1.93	&	5.2,4.2	\\
20090121$^*$	&	I	&	30	&	2.51	&	5.2	\\
20090122$^*$	&	R,I	&	6,13	&	0.81,1.43	&	5.2,4.2	\\
20090128	&	V,R,I	&	10,10,10	&	0.12	&	0.3	\\
20090322$^*$	&	B,V,R	&	5,5,5	&	0.06,0.04,0.07	&	0.7,0.6,0.5	\\
20090323$^{*\dag}$	&	B,V,R	&	5,5,5	&	0.14	&	2	\\
20090326$^{*\dag}$	&	B,V,R	&	5,5,5	&	0.13	&	1.9	\\
20090327$^{*\dag}$	&	B,V,R	&	5,5,5	&	0.25,0.24,0.23	&	3.7,3.5,3.4	\\
20091014$^{*\dag}$	&	B,V,R,I	&	10,10,10	&	0.19	&	1.2	\\
20091021$^\dag$	&	B,V,R,I	&	10,10,10,10	&	0.2	&	1.3	\\
20091024$^*$	&	V,R,I	&	5,5,5	&	0.27,0.21,0.16	&	4,3,2.3	\\
20091026$^{*\dag}$	&	B,V,R,I	&	10,10,10,10	&	0.24	&	1.5	\\
20091027$^\dag$	&	B,V,R,I	&	17,17,17,17	&	0.43	&	1.5	\\
20091028$^\dag$	&	B,V,R,I	&	25,25,25,25	&	0.64	&	1.5	\\
20091030$^\dag$	&	B,V,R,I	&	10,10,10,10	&	0.24	&	1.5	\\
20091106$^{*\dag}$	&	B,V,R,I	&	31,34,160,31	&	0.77,0.79,1.09,0.77	&	1.5,1.4,0.4,1.5	\\
20091107$^{*\dag}$	&	B,V,R,I	&	33,33,33,34	&	0.58	&	1	\\
20091108$^{*\dag}$	&	B,V,R,I	&	25,25,25,25	&	0.41	&	1	\\
20091110$^{*\dag}$	&	B,V,R,I	&	9,9,9,9	&	0.15	&	1	\\
20091113$^*$	&	R,I	&	4,5	&	0.32,0.25	&	6.5,3.7	\\
20091114$^*$	&	V,R,I	&	5,5,5	&	0.39,0.24,0.18	&	5.8,3.5,2.7	\\
20091115$^*$	&	V,R,I	&	1,5,5	&	0.3,0.18	&	4.4,2.6	\\
20091214$^{*\dag}$	&	B,V,R,I	&	20,20,20,20	&	0.51	&	1.6	\\
20100104$^{*\dag}$	&	V,R,I	&	10,10,10	&	0.76,0.78,0.72	&	5,5.1,4.7	\\
20100106$^*$	&	V,R,I	&	9,15,10	&	0.04,0.15,0.03	&	0.3,0.22,0.22	\\
20100109$^*$	&	V,R,I	&	25,25,35	&	3.11,3.06,3.88	&	3.5,2.6,1.8	\\
20100110$^*$	&	V,R,I	&	5,5,5	&	0.45,0.24,0.21	&	4.3,3.6,3.2	\\
20100111$^*$	&	I	&	23	&	1.37	&	3.7	\\
20100114$^*$	&	V,R,I	&	5,5,5	&	0.34,0.23,0.17	&	5.1,3.4,2.5	\\
20100116$^*$	&	R	&	144	&	0.97	&	0.22	\\
20100215$^{*\dag}$	&	B,V,R,I	&	10,10,10,9	&	0.41	&	2.7	\\
20100220$^*$	&	V,R,I	&	5,5,5	&	0.32,0.28,0.16	&	3.7,3.3,2	\\
20100221$^*$	&	V,R,I	&	5,5,5	&	0.15,0.13,0.11	&	2.1,1.9,1.6	\\
20100222$^{*\dag}$	&	B,V,R,I	&	32,32,32,32	&	0.79	&	1.5	\\
20100226$^*$	&	B,V,R,I	&	42,42,42,42	&	0.72,0.6,0.5,0.79	&	0.38,0.3,0.2,0.2	\\
20100227$^*$	&	B,V,R,I	&	10,10,10,10	&	0.06,0.05,0.03,0.03	&	0.3,0.3,0.2,0.2	\\
20100228$^*$	&	B,V,R,I	&	20,20,20,20	&	0.26,0.25.0.23,0.25	&	0.3,0.3,0.2,0.2	\\
20100311$^*$	&	B,V,R,I	&	20,20,20,20	&	2.45,2.44,2.43,2.23	&	0.3,0.3,0.2,0.2	\\
20100319$^*$	&	R,I	&	5,65	&	0.18,2.59	&	2.6,2.2	\\
20100407$^*$	&	V,R,I	&	15,15,14	&	0.13,0.11,0.1	&	0.5,0.4,0.4	\\
20100408$^*$	&	V,R,I	&	10,10,10	&	0.17,0.26,0.07	&	0.5,0.4,0.4	\\
20110130$^\dag$	&	V,R,I	&	5,5,5	&	0.18	&	2.7	\\
20111112$^\dag$	&	V,R,I	&	5,5,5	&	0.59	&	8.9	\\
20111113$^\dag$	&	V,R,I	&	5,5,5	&	0.76	&	11.4	\\
20111114$^\dag$	&	V,R,I	&	5,5,5	&	0.48	&	7.2	\\
20111115$^\dag$	&	V,R,I	&	5,5,5	&	0.59	&	8.9	\\
20111116$^\dag$	&	V,R,I	&	5,5,5	&	0.48	&	7.2	\\
20120224$^\dag$	&	B,R	&	11,11	&	0.62	&	3.4	\\
20120225$^\dag$	&	B,I	&	20,20	&	0.78,0.85	&	2.4,2.6	\\
20120226$^\dag$	&	B,I	&	28,31	&	1.22	&	2.4	\\
20120227$^\dag$	&	V,R,I	&	4,5,5	&	0.33	&	4.9	\\
20120228$^\dag$	&	V,R,I	&	5,5,5	&	0.38	&	5.7	\\
20120229$^\dag$	&	V,R,I	&	5,5,5	&	0.38	&	5.7	\\
20120401$^\dag$	&	V,R,I	&	9,10,10	&	0.85	&	5.6	\\
20120402$^\dag$	&	V,R,I	&	5,5,5	&	0.26	&	3.9	\\
20120403$^\dag$	&	V,R,I	&	5,5,5	&	0.29	&	4.3	\\
20121224$^\dag$	&	B,V,R,I	&	4,3,32,4	&	0.37,0.24,1.22,0.36	&	7.3,7.3,2.3,7.3	\\
\enddata
\tablecomments{The ``$^\dag$'' stands for measurements in the first mode and the ``$^*$'' stands for that these dates are same as that in Liao et al. (2014).
}
\end{deluxetable}

\begin{deluxetable}{cccr}
\tablecaption{The Magnitude of S5 0716+714 ($B$ Filter)\label{tbl-2}} \tablewidth{0pt}
\tablehead{\colhead{Date} & \colhead{MJD} & \colhead{Magnitude}
& \colhead{$\sigma$}} \startdata
2005 Apr 05 	&	53465.918 	&	14.061 	&	0.013 	\\
2005 Apr 05 	&	53465.928 	&	14.084 	&	0.013 	\\
2005 Apr 06 	&	53466.844 	&	14.266 	&	0.020 	\\
2005 Apr 06 	&	53466.855 	&	14.318 	&	0.020 	\\
2005 Apr 06 	&	53466.864 	&	14.327 	&	0.020 	\\

\enddata
\tablecomments{The first, second, third,
and fourth columns are the universal time of observation, the corresponding modified Julian day (${\rm MJD}={\rm JD}-2400000.5$), the magnitude and the rms error, respectively. This table is published in its entirety in the machine-readable form.}
\end{deluxetable}

\begin{deluxetable}{cccr}
\tablecaption{The Magnitude of S5 0716+714 ($V$ Filter)\label{tbl-2}} \tablewidth{0pt}
\tablehead{\colhead{Date} & \colhead{MJD} & \colhead{Magnitude}
& \colhead{$\sigma$}} \startdata
2005 Apr 05 	&	53465.916 	&	13.691 	&	0.002 	\\
2005 Apr 05 	&	53465.926 	&	13.659 	&	0.002 	\\
2005 Apr 06 	&	53466.841 	&	13.857 	&	0.016 	\\
2005 Apr 06 	&	53466.852 	&	13.873 	&	0.016 	\\
2005 Apr 06 	&	53466.862 	&	13.899 	&	0.016 	\\

\enddata
\tablecomments{The first, second, third,
and fourth columns are the universal time of observation, the corresponding modified Julian day (${\rm MJD}={\rm JD}-2400000.5$), the magnitude and the rms error, respectively. This table is published in its entirety in the machine-readable form.}
\end{deluxetable}

\begin{deluxetable}{cccr}
\tablecaption{The Magnitude of S5 0716+714 ($R$ Filter)\label{tbl-3}} \tablewidth{0pt}
\tablehead{\colhead{Date} & \colhead{MJD} & \colhead{Magnitude}
& \colhead{$\sigma$}} \startdata
2005 Apr 05 	&	53465.908 	&	13.175 	&	0.035 	\\
2005 Apr 05 	&	53465.924 	&	13.149 	&	0.035 	\\
2005 Apr 06 	&	53466.839 	&	13.496 	&	0.010 	\\
2005 Apr 06 	&	53466.850 	&	13.480 	&	0.010 	\\
2005 Apr 06 	&	53466.860 	&	13.476 	&	0.010 	\\

\enddata
\tablecomments{The first, second, third,
and fourth columns are the universal time of observation, the corresponding modified Julian day (${\rm MJD}={\rm JD}-2400000.5$), the magnitude and the rms error, respectively. This table is published in its entirety in the machine-readable form.}
\end{deluxetable}

\begin{deluxetable}{cccr}
\tablecaption{The Magnitude of S5 0716+714 ($I$ Filter)\label{tbl-4}} \tablewidth{0pt}
\tablehead{\colhead{Date} & \colhead{MJD} & \colhead{Magnitude}
& \colhead{$\sigma$}} \startdata
2005 Apr 05 	&	53465.905 	&	12.890 	&	0.008 	\\
2005 Apr 05 	&	53465.922 	&	12.867 	&	0.008 	\\
2005 Apr 06 	&	53466.836 	&	13.021 	&	0.015 	\\
2005 Apr 06 	&	53466.847 	&	13.003 	&	0.015 	\\
2005 Apr 06 	&	53466.867 	&	13.025 	&	0.015 	\\
\enddata
\tablecomments{The first, second, third,
and fourth columns are the universal time of observation, the corresponding modified Julian day (${\rm MJD}={\rm JD}-2400000.5$), the magnitude and the rms error, respectively. This table is published in its entirety in the machine-readable form.}
\end{deluxetable}

\begin{deluxetable}{ccccccccr}
\small
\tablecaption{Statistic Results for IDV\label{tbl-5}}
\tablewidth{0pt} \tablehead{\colhead{Date} & \colhead{Bands}  & \colhead{Number
} & \colhead{$F$} &\colhead{$F_C(99)$}
&\colhead{$F_A$} &\colhead{$F_A(99)$} &\colhead{V/N} &\colhead{A(\%)}} \startdata
20061028	&	I	&	21	&	1.41 	&	2.94 	&	8.07 	&	5.19 	&	PV	&	 	\\
20061128	&	V	&	32	&	0.70 	&	2.35 	&	7.31 	&	3.82 	&	PV	&		\\
20061128	&	I	&	31	&	1.71 	&	2.39 	&	2.27 	&	3.86 	&	N	&	 	\\
20070128	&	I	&	16	&	2.07 	&	3.52 	&	1.48 	&	6.70 	&	N	&		\\
20070327	&	V	&	20	&	0.71 	&	3.03 	&	1.27 	&	5.29 	&	N	&		\\
20070327	&	I	&	20	&	1.22 	&	3.03 	&	3.81 	&	5.29 	&	N	&		\\
20070328	&	B	&	15	&	0.80 	&	3.70 	&	9.85 	&	6.93 	&	PV	&		\\
20070328	&	V	&	15	&	1.67 	&	3.70 	&	25.5 	&	6.93 	&	PV	&		\\
20070328	&	I	&	15	&	1.68 	&	3.70 	&	4.43 	&	6.93 	&	N	&	 	\\
20070416	&	I	&	15	&	0.67 	&	3.70 	&	0.44 	&	6.93 	&	N	&	 	\\
20080110	&	I	&	20	&	1.14 	&	3.03 	&	7.24 	&	5.29 	&	PV	&	 	\\
20080111	&	I	&	20	&	4.17 	&	3.03 	&	21.55 	&	5.29 	&	V	&	2.99 	\\
20081002	&	I	&	23	&	1.52 	&	2.78 	&	2.02 	&	5.01 	&	N	&	 	\\
20090116	&	I	&	20	&	8.53 	&	3.03 	&	54.63 	&	5.29 	&	V	&	4.61 	\\
20090117	&	I	&	15	&	6.16 	&	3.70 	&	62.80 	&	6.93 	&	V	&	4.69 	\\
20090120	&	I	&	15	&	1.18 	&	3.70 	&	34.27 	&	6.93 	&	PV	&		\\
20090120	&	R	&	15	&	0.76 	&	3.70 	&	3.08 	&	6.93 	&	N	&	 	\\
20090121	&	I	&	30	&	1.45 	&	2.42 	&	5.30 	&	3.90 	&	PV	&	 	\\
20091027	&	B	&	17	&	9.74 	&	3.37 	&	61.54 	&	6.51 	&	V	&	2.77 	\\
20091027	&	V	&	17	&	8.57 	&	3.37 	&	33.84 	&	6.51 	&	V	&	2.67 	\\
20091027	&	R	&	17	&	6.80 	&	3.37 	&	23.27 	&	6.51 	&	V	&	2.61 	\\
20091027	&	I	&	17	&	15.75 	&	3.37 	&	24.11 	&	6.51 	&	V	&	3.13 	\\
20091028	&	B	&	25	&	4.89 	&	2.66 	&	5.55 	&	4.43 	&	V	&	1.62 	\\
20091028	&	V	&	25	&	3.39 	&	2.66 	&	29.31 	&	4.43 	&	V	&	1.55 	\\
20091028	&	R	&	25	&	1.11 	&	2.66 	&	11.42 	&	4.43 	&	PV	&		\\
20091028	&	I	&	25	&	1.50 	&	2.66 	&	5.24 	&	4.43 	&	PV	&	 	\\
20091106	&	B	&	31	&	0.50 	&	2.39 	&	6.33 	&	3.86 	&	PV	&	 	\\
20091106	&	V	&	34	&	0.45 	&	2.29 	&	5.38 	&	3.75 	&	PV	&	 	\\
20091106	&	R	&	160	&	1.05 	&	1.45 	&	5.97 	&	1.84 	&	PV	&	 	\\
20091106	&	I	&	31	&	0.70 	&	2.39 	&	11.03 	&	3.86 	&	PV	&		\\
20091107	&	B	&	33	&	0.97 	&	2.32 	&	1.18 	&	3.78 	&	N	&		\\
20091107	&	V	&	33	&	0.88 	&	2.32 	&	1.80 	&	3.78 	&	N	&		\\
20091107	&	R	&	33	&	0.99 	&	2.32 	&	0.74 	&	3.78 	&	N	&		\\
20091107	&	I	&	34	&	0.91 	&	2.29 	&	0.81 	&	3.75 	&	N	&		\\
20091108	&	B	&	25	&	0.83 	&	2.66 	&	1.73 	&	4.43 	&	N	&		\\
20091108	&	V	&	25	&	0.87 	&	2.66 	&	1.11 	&	4.43 	&	N	&		\\
20091108	&	R	&	25	&	0.75 	&	2.66 	&	2.03 	&	4.43 	&	N	&		\\
20091108	&	I	&	25	&	0.74 	&	2.66 	&	2.01 	&	4.43 	&	N	&		\\
20091214	&	B	&	19	&	1.27 	&	3.13 	&	0.17 	&	6.23 	&	N	&		\\
20091214	&	V	&	20	&	1.66 	&	3.03 	&	0.72 	&	5.29 	&	N	&		\\
20091214	&	R	&	20	&	1.80 	&	3.03 	&	0.38 	&	5.29 	&	N	&		\\
20091214	&	I	&	20	&	0.94 	&	3.03 	&	3.77 	&	5.29 	&	N	&		\\
20100106	&	R	&	15	&	1.57 	&	3.70 	&	2.49 	&	6.93 	&	N	&		\\
20100109	&	V	&	25	&	2.94 	&	2.66 	&	27.64 	&	4.43 	&	V	&	6.16 	\\
20100109	&	I	&	35	&	5.32 	&	2.26 	&	56.28 	&	3.53 	&	V	&	4.26 	\\
20100109	&	R	&	25	&	1.87 	&	2.66 	&	20.71 	&	4.43 	&	PV	&	 	\\
20100111	&	I	&	20	&	0.93 	&	2.86 	&	0.39 	&	5.29 	&	N	&		\\
20100116	&	R	&	144	&	9.91 	&	1.48 	&	63.73 	&	1.91 	&	V	&	4.12 	\\
20100222	&	B	&	32	&	1.90 	&	2.35 	&	1.71 	&	3.82 	&	N	&		\\
20100222	&	V	&	32	&	1.51 	&	2.35 	&	2.61 	&	3.82 	&	N	&		\\
20100222	&	R	&	32	&	1.31 	&	2.35 	&	2.12 	&	3.82 	&	N	&		\\
20100222	&	I	&	32	&	1.92 	&	2.35 	&	2.61 	&	3.82 	&	N	&	 	\\
20100226	&	B	&	42	&	14.27 	&	2.10 	&	45.38 	&	2.54 	&	V	&	5.96 	\\
20100226	&	V	&	42	&	18.06 	&	2.10 	&	40.05 	&	2.54 	&	V	&	6.36 	\\
20100226	&	R	&	41	&	13.37 	&	2.10 	&	7.45 	&	2.54 	&	V	&	4.34 	\\
20100226	&	I	&	42	&	6.75 	&	2.10 	&	12.14 	&	2.54 	&	V	&	5.21 	\\
20100228	&	B	&	20	&	0.49 	&	3.03 	&	5.11 	&	5.29 	&	N	&		\\
20100228	&	V	&	20	&	0.92 	&	3.03 	&	0.49 	&	5.29 	&	N	&		\\
20100228	&	R	&	20	&	0.63 	&	3.03 	&	1.35 	&	5.29 	&	N	&		\\
20100228	&	I	&	20	&	0.53 	&	3.03 	&	0.58 	&	5.29 	&	N	&		\\
20100311	&	B	&	20	&	1.59 	&	3.03 	&	3.97 	&	5.29 	&	N	&		\\
20100311	&	V	&	20	&	0.83 	&	3.03 	&	12.75 	&	5.29 	&	PV	&		\\
20100311	&	R	&	20	&	1.48 	&	3.03 	&	1.78 	&	5.29 	&	N	&		\\
20100311	&	I	&	19	&	2.24 	&	3.13 	&	72.62 	&	6.23 	&	PV	&		\\
20100319	&	I	&	65	&	6.12 	&	1.80 	&	22.20 	&	2.55 	&	V	&	4.29 	\\
20100407	&	V	&	15	&	1.38 	&	3.70 	&	3.70 	&	10.87 	&	N	&		\\
20100407	&	R	&	15	&	0.95 	&	3.70 	&	3.70 	&	5.74 	&	N	&		\\
20120225	&	B	&	20	&	0.72 	&	3.03 	&	0.54 	&	5.29 	&	N	&		\\
20120225	&	I	&	20	&	1.28 	&	3.03 	&	6.93 	&	5.29 	&	PV	&	 	\\
20120226	&	B	&	28	&	0.67 	&	2.51 	&	2.99 	&	4.26 	&	N	&	 	\\
20120226	&	I	&	31	&	0.49 	&	2.39 	&	3.58 	&	3.86 	&	N	&	 	\\
20121224	&	R	&	32	&	2.11 	&	2.35 	&	0.63 	&	3.82 	&	N	&		\\
\enddata
\tablecomments{Column 4 and column 5 are the average $F$ value and the
critical $F$ value with 99 per cent confidence level respectively. Column 6 and column 7 are the
$F$ value of ANOVA and the critical $F$ value with 99
per cent confidence level (${\rm bin}=5$) respectively. Column 8 and column 9 are the variability status (V:
variable, N: non-variable, PV: probable variable) and the
variability amplitude (Heidt \& Wagner 1996) respectively.}
\end{deluxetable}

\end{document}